\documentclass[prd,aps,floats,epsf,twocolumn,nofootinbib]{revtex4}
\usepackage{graphicx}

\newcommand{\be}{\begin{equation}}
\newcommand{\ee}{\end{equation}}
\newcommand{\ba}{\begin{eqnarray}}
\newcommand{\ea}{\end{eqnarray}}

\begin{document}
\title{\bf Anisotropic brane cosmologies with exponential potentials}
\author{J.M. Aguirregabiria and Ruth Lazkoz
\\
{\it   Fisika Teorikoa,
Euskal Herriko Unibertsitatea}\\
{\it 644 Posta Kutxatila, 48080 Bilbao, Spain}\\
{\it wtpagagj@lg.ehu.es}\\
{\it wtplasar@lg.ehu.es}}

\begin{abstract}
We study Bianchi I type brane cosmologies 
with scalar matter self-interacting through combinations of  exponential potentials. 
Such models correspond in some cases to inflationary universes. We discuss in detail
the conditions for accelerated expansion to occur: in particular, we show that the condition which is necessary and sufficient for inflation in the relativistic version of the models
is not sufficient in the brane case. Another peculiar feature of the models is that
the relationship between the value of the scale factor at the beginning of inflation and the equation of state is very different from what one finds  in the relativistic framework.
We also analyze the influence of the value of the anisotropy and the brane tension, and show that the associated effects 
become negligible in the late time limit, those
related to the anisotropy disappearing earlier. This study focuses mainly
on single field models, but we also consider a generalization yielding models
with multiple non-interacting fields and 
examine their features  briefly. We conclude that, in the brane scenario,  an increase
in the number of fields assists inflation, as happens
in general relativity.
\end{abstract}
\maketitle

\section{Introduction}

According to the most promising candidates for a quantum gravity theory,
we are living in more than four dimensions \cite{Pol98}, 
and gravity is  effectively four dimensional only at low enough energies. 
This has inspired scenarios
in cosmology different from  the standard one. 
The  ``brane world'' picture is the most recent of them 
all \cite{RanSun99}, and assumes that ordinary matter is confined to a 
hypersurface with three spatial dimensions
embedded in a multidimensional space-time, allowing gravity act in
the fifth dimension.

The ``brane-world'' proposal has motivated a revision of the predictions
to be tested by future observations. Even if in a first approach isotropy is assumed, one will immediately find  \cite{BinDefLan00} that the evolution of the Universe in this new scenario is rather peculiar, at least at early times, because of the modifications arising in the Hubble equation. Given this, one may wonder for instance what aspects of cosmological inflation should be changed to fit in this alternative description. 
Clearly, one of the most important problems to 
address within this framework
is that of the initial conditions.
Recently, it has been suggested \cite{MaaSahSai01} that
anisotropic spacetimes typically expand faster than
isotropic ones. This study was grounded on a
qualitative analysis of the evolution equations for the scalar field 
(Klein-Gordon equation) and the average expansion rate  
(generalized Raychaudhuri equation) for Bianchi I type cosmologies.

In this spirit, we have investigated the same equation set, together with its solutions, 
just assuming at the beginning that the potential depends on the scale factor
in such a way that it is compatible with inflation. In particular, we have
found that, under some assumptions, the matter source can be cast in the form of
a  scalar field self-interacting through a combination of exponential potentials.
Such models were first considered in the context of inflation nearly two decades ago
\cite{LucMat85}, and have been paid much attention along the years because of 
several reasons. First of all, there is solid motivation 
that makes them physically appealing, for they appear in four-dimensional
effective Kaluza-Klein theories arising from compactification
of higher dimensional supergravity or superstring theories (see \cite{Tow01}
for a review).
In addition, it was shown long ago that FRW power law models, 
which are
driven by exponential potentials,  behave as attractors in the sense that
models with broad disparity in their initial conditions become power law 
solutions at late times \cite{Hal87,PebRat88}. Last but not least, exponential 
potentials have proved very useful in general relativity 
for providing new exact solutions because
they add a rather small degree of
non-linearity to the field equations \cite{AguFeiIba93}.
Unfortunately, the equations one has to solve in brane cosmology are 
intrinsically so non-linear that, even if one considers  just exponential 
potentials, the task of finding exact solutions is a hard one,  except in very 
specific cases.

With the aim of getting round this difficulty, and as mentioned before,
we will not assume from the start that
we have an exponential potential or a combination of such terms, 
but rather end up having it after making a certain ansatz regarding
time evolution.

As expected, if the expansion is fast enough then anisotropy exerts less
influence on the expansion than the quadratic energy-momentum corrections due
to the extrinsic curvature. It follows then that anisotropy 
does not prevent inflation, as
one  also gets in the framework of general relativity \cite{MosSah86,
TurWid86,JenSte86}.  In the late time regime, the effect of anisotropy and extra dimensions are negligible, and the solutions behave as if they were isotropic 
and purely relativistic.  

The qualitative behavior of Bianchi I type brane cosmologies 
has been studied using dynamical systems techniques by several authors.
Barotropic perfect fluids were considered by Campos and Sopuerta in the absence \cite{CamSop01a}
and presence \cite{CamSop01b} of bulk effects; whereas 
Goheer and Dunsby \cite{GohDun03} considered models with a single scalar field (see also  \cite{GohDun02} for an earlier
study along the same lines focusing on isotropic cosmologies). According to those studies, late time isotropization is an inherent feature to the models. This idea of inflation acting as an isotropizing agent, that
was put forward years ago by Wald \cite{Wal82}, has been recently revisited in the brane scenario \cite{SanVerFer01} (see also \cite{IbaHooCol95,ColHooIba97} for a discussion on the isotropization
of anisotropic relativistic cosmologies with an exponential potential). Nevertheless, we have recently shown \cite{AguChiLaz03} 
that the brane might interact with the bulk in a way not compatible with late time isotropization. Here, however, we will discard the bulk effects and 
Wald's argument will still apply.

In another recent study \cite{CheHarMak01}, Bianchi I type
exact solutions in parametric form were found 
for barotropic perfect fluids and several scalar matter models. 
It was shown that for exponential potentials the simple relation 
$\phi\propto \log t$ will not hold at all times. This does not exclude, 
though, the possibility of having this behavior at some epoch, 
and that is indeed the case for the late time regime of 
our solutions. We also show here that the effects of the extra dimensions make
the models isotropize faster, this being linked to an increase in the Hubble
rate.

Another interesting aspect of the problem is that, typically, exponential 
potentials only render inflation if their slope is small enough. Since our solutions mimic their relativistic counterparts late in their 
history, it is not surprising that the same restrictions on the values 
of the slope will hold for our models too. This is discussed in Section 3.
However, one could  still have inflation with sloppy potentials 
if there are enough non-interacting fields. This is the so called
assisted inflation proposal. We also analyze this possibility in the last section, 
and find that  inflation is likelier in such multifield models because the Hubble factor grows as the number of fields increases.

The plan of the paper is as follows: in Sec. 2 we study the equations of 
motion for the scalar field and the gravitational 
field equations discarding bulk effects and under the assumption
of a Bianchi type I geometry. In Sec. 3 we consider solutions such that their 
potential energy decreases as the models expand,  and find approximate solutions
with power law inflation at very late times. Sec. 4  is  devoted to  discussing in detail the inflationary behavior, paying special 
attention to the role of  anisotropy and the corrections from extra dimensions. In this 
section, we show that the value of the expansion factor at the beginning of inflation depends on the equation of state in an peculiar way,
 if we compare the results with what one gets in the 
relativistic. This suggests that the evolution of brane cosmologies is in some aspects
very different from what is regarded as standard, and that the subject deserves further investigation.  In Sec. 5 we turn to consider multi-field generalizations 
and examine their features. We show that it is possible to model assisted inflation in the 
brane setup too. Finally, in Sec. 6 we draw our main 
conclusions, and review future prospects.

\section{Geometry on the brane and field equations}
We assume that the 5-dimensional line element
giving the geometry of the bulk is of the form \cite{ShiMaeSas00}
\be
d\tilde s^2=\tilde g_{AB}dx^{A}dx^{B},
\ee
where $A,B=0,1,2,3,4$. Tildes will be used to denote the bulk 
counterparts of standard general relativistic quantities.
If we now denote by $n^{A}$ the unit vector normal to the brane, then 
the metric induced on that hypersurface is
$g_{AB}=\tilde g_{AB}-n_An_B$. It is convenient to define
a new coordinate $\chi$ such that the brane is located at $\chi=0$ and
$n_A dx^A=d\chi$. This renders the line element in the alternative form
 \be
ds^2=d\chi^2+g_{\mu\nu}dx^{\mu}dx^{\nu}
\ee
with $\mu,\nu=0,1,2,3$.  The five-dimensional
bulk field equations are
\be
\tilde G_{AB}=-{\tilde \kappa} ^2[\tilde \Lambda \tilde g_{AB}+
\delta(\chi)(\lambda g_{AB}-T_{AB})].
\ee

Effectively, the Einstein equations on the brane take the form
\be
G_{\mu\nu}=-\Lambda g_{\mu \nu}+\kappa^2T_{\mu\nu}+
\tilde\kappa^4 S_{\mu\nu}-E_{\mu\nu},
\ee
where  
\be
S_{\mu\nu}=\frac{1}{12}TT_{\mu\nu}-
\frac{1}{4}{T_{\mu}}^{\alpha}T_{\nu\alpha}+
\frac{1}{24}g_{\mu\nu}\left(3T^{\alpha\beta}T_{\alpha\beta}-T^2\right),
\ee
$ \Lambda=\tilde \kappa^2(\tilde\Lambda+\tilde \kappa^2\lambda^2/6)/2$,
$\kappa^2=\tilde\kappa^4\lambda/6$, $T=T_{\alpha}^{\alpha}$,
 and $E_{AB}=C_{AIBJ}n^In^J$. 
Here $C_{AIBJ}$ is the five dimensional Weyl tensor in the bulk, whereas
$T_{\mu\nu}$ and $\lambda$ stand for the values on the brane of the
energy-momentum tensor and the vacuum
energy respectively. The equations were obtained under the assumption
of $Z_2$ symmetry, and confinement of the matter fields to the brane,
in agreement with the brane scenario devised by Horava and Witten 
\cite{HorWit96}.

It has been shown that $E_{\mu\nu}$ can be decomposed like
\begin{equation}
E_{\mu\nu}=-\left(\frac{\tilde \kappa}{\kappa}\right)^4\left[{\cal U}\left(u_\mu u_\nu+\frac{1}{3}\,h_{\mu\nu}\right)+{\cal P}_{\mu\nu}+2{\cal Q}_{(\mu}u_{\nu)}\right],
\end{equation}
where $u^{\mu}$ is  the four velocity of an observer on
the brane, ${\cal U}$ is the effective non local energy density on the brane, ${\cal P}_{\mu\nu}$
is the effective nonlocal anisotropic stress, and 
 $Q_\mu$ is an effective non local energy flux on the brane, which vanishes identically
 for a Bianchi type I brane. In addition ${\cal P}_{\mu\nu}$ and $E_{\mu\nu}$
 are subject to the constraints
 \begin{eqnarray}&&D^{\nu}{\cal P}_{\mu\nu}=0\label{consp},\\
 &&\nabla^{\mu}E_{\mu\nu}={\tilde \kappa}^4\nabla^{\mu}S_{\mu\nu}\label{consu}.
 \end{eqnarray}
 
Let us consider now four-dimensional Bianchi I type spacetimes. The line
element is given by
\begin{equation}
ds ^2=-dt^2+a_1^2(t)dx^2+a_2^2(t)dy^2+a_3^2(t)dz^2.
\end{equation}

As usual, we define the expansion rates along 
the three spatial directions  as $H_i=\dot a_i/a_i$ for $i=1,2,3$.  The shear tensor will be denoted as $\sigma_{\mu\nu}$ and 
\begin{equation}
\sigma^{\mu\nu}\sigma_{\mu\nu}=\sum_{i=1}^3(H_i-H)^2,
\end{equation}
where $3H=H_1+H_2+H_3$.
For this metric, Eq. (\ref{consp}) is  identically satisfied. Moreover,
since there is no evolution equation for ${\cal P}_{\mu\nu}$ on the brane, one usually assumes that it is either null or satisfies the less restrictive condition that
 $\sigma ^{\mu\nu} {\cal P}_{\mu\nu}=0$ (It has been proved \cite{ColHooYe02} that the integrability conditions
for ${\cal Q}_{\mu}=0$, ${\cal P}_{\mu\nu}=0$ imply spatial homogeneity, so this is consistent with having a Bianchi I metric on the brane.). Both assumptions on ${\cal P}_{\mu\nu}$ 
transform Eq. (\ref{consu}) into  a very simple evolution equation for ${\cal U}$, namely
\begin{equation}
\dot{\cal U}+4 H{\cal U}=0. 
\end{equation}
The solution of the latter corresponds to non local energy evolving like radiation, 
i.e. ${\cal U}={{\cal U}_{\,0}}/a^4$, with ${{\cal U}_{\,0}}$ a 
constant that can be null. As we see,  
the joint choice of ${\cal P}_{\mu\nu}=0$ and ${\cal U}=0$ is consistent on the brane. 
Whether this is also fully consistent for the bulk  remains an open question, but we follow other authors in making these arguably strong assumptions \cite{MaaSahSai01,SolDunEll01}. This  allows gaining considerable insight in the changes on the evolution
of cosmological models associated with the modification of the Friedmann equation,
which might be otherwise concealed by bulk effects.


 Under the assumptions made, the
gravitational field equations and the Klein-Gordon equation read
\begin{equation} \label{BI1b}
H_1H_2+H_1H_3+H_2H_3=\Lambda  +
{\kappa^2}\rho+ 
\frac{\tilde\kappa^4}{12}\rho^2,
\end{equation}
\begin{equation} \label{BI2b}
H_2^2+H_2H_3+H_3^2+\dot H_2+ \dot H_3=
\Lambda -{{{\kappa }}}^2\, p  - 
\frac{\rho}{12}{\left( 2\,p + \rho  \right) \,{{{\tilde\kappa }}}^4},
\end{equation}
\begin{equation} \label{BI3b}
H_1^2+H_1H_3+H_3^2+\dot H_1+ \dot H_3=\Lambda -{{{\kappa }}}^2\, p  - 
\frac{\rho}{12}{\left( 2\,p + \rho  \right) \,{{{\tilde\kappa }}}^4},
\end{equation}
\begin{equation} \label{BI4b}
H_1^2+H_1H_2+H_2^2+\dot H_1+ \dot H_2=\Lambda -{{{\kappa }}}^2\, p  - 
\frac{\rho}{12}{\left( 2\,p + \rho  \right) \,{{{\tilde\kappa }}}^4},
\end{equation}
\begin{equation} \label{BI5}
\ddot \phi+3H\dot \phi+\frac{\partial V}{\partial \phi } =0,
\end{equation}
where we have set
\begin{eqnarray}
\rho=\frac{1}{2}\dot \phi^2+V(\phi),\\
p=\frac{1}{2}\dot \phi^2-V(\phi),
\end{eqnarray}
according to the customary interpretation of irrotational scalar fields
in terms of a perfect fluid.

 Due to the isotropy of the matter source, 
one can obtain constraint equations which provide
a relationship expressing the expansion
rate along one direction in terms of the other two and their derivatives.

Even though the solutions to the system (\ref{BI1b})--(\ref{BI4b})
include locally rotationally symmetric models (LRS), i.e., models
with the same expansion rate along two directions,
we will not consider them any further. Thus, 
in what follows we will assume $H_1\neq H_2$, $H_2\neq H_3$, $H_3\neq H_1$,
so that the models admit only three isometries.  The constraints arising 
from having identical pressure along the three spatial directions are

\begin{equation} \label{c2}
H_1+H_2+H_3=-\frac{\dot H_2-\dot H_1}{H_2-H_1}
=-\frac{\dot H_3-\dot H_2}{H_3-H_2}
=-\frac{\dot H_1-\dot H_3}{H_1-H_3}.
\end{equation}

This allows to distinguish two different cases. The first one corresponds 
to $\dot H_2\neq \dot H_3$. Hence $\dot H_1\neq \dot H_2$ and $\dot 
H_3\neq
\dot H_1$.
Integration of Eq. (\ref{c2}) yields the following relationships:
\begin{equation} \label{ca}
H_1-H_3=\vartheta(H_1-H_2)= \frac{\vartheta}{\vartheta -1}(H_2-H_3),
\end{equation}
where $\vartheta\ne 1$ is an arbitrary integration constant. The second case
arises when the Hubble factors along any two directions differ only by a 
constant, then necessarily $H_1+H_2+H_3=0$ and we get a model which does not 
evolve on average. For simplicity we will discard these cases too.

It is convenient to write the field equations using the following definitions:
\begin{eqnarray}
&&P\equiv H_1-H_2,\label{P}\\
&&Q\equiv H_1-H_3\label{Q}.
\end{eqnarray}
We will also use the following constant:
\be{\varpi}\equiv\frac{1-\vartheta+\vartheta^2}{3}\ge \frac{1}{4}\label{varpi}.\ee
Note that, since we are not considering LRS models, we will have
${\varpi}\ne 1/3$. Equations (\ref{BI1b})--(\ref{BI5}) become
\begin{eqnarray}
&&Q=\vartheta P\label{QP}, \\
&&H=-\frac{\dot P}{3P}\label{defH},\\
&&\Lambda  +
{\kappa^2}\rho+ 
\frac{\tilde\kappa^4}{12}\rho^2
=3H^2-{\varpi} P^2, \label{friedmann}\\
&&\ddot \phi+3H\dot \phi+\frac{\partial V}{\partial \phi } =0.\label{KG}
\end{eqnarray}

\section{Cosmologies with exponential potentials}
In this section we look for  solutions 
by means of an unconventional technique that consists in taking the average scale factor
$a=(a_1a_2a_3)^{1/3}$ as the independent variable, instead of the time variable $t$. 
This procedure has been successfully used in related problems. 

Taking advantage of the insight gained in Ref. \cite{ChiJak96}, we write
\be
V(\phi(a))=\frac{F(a)}{a^6}
\ee
and make the change of variables $dt=a^3d\eta$ in Eq. (\ref{KG}), so that
\be
\frac{d^2\phi}{d\eta^2}+a^6\frac{dV}{d\phi}=0,
\ee
which in turn gives us the first integral
\be
\frac{1}{2}\dot \phi^2+V(\phi)-\frac{6}{a^6}\int \frac{F}{a}da=\frac{c}{a^6},
\ee
where $c$ is an arbitrary integration constant. This is equivalent
to saying that
\be
\rho(a)=\frac{6}{a^6}\int \frac{F}{a}da+\frac{c}{a^6}.
\ee

Combining $H=\dot a/a$ with Eq. (\ref{defH}) we 
get $P=c' a ^{-3}$, where $c'$ is another arbitrary integration constant
that we will set to one for simplicity.
Now, using Eqs. (\ref{defH}) and (\ref{friedmann}) we can formally reduce the 
problem to 
quadratures, namely 
\be
t=t_0\pm\int \frac{\sqrt{3}\,da}{a
\sqrt{\Lambda+\kappa^2\rho(a)+\displaystyle\frac{\tilde\kappa^4}{12}
\,\rho(a)^2+\displaystyle\frac{{\varpi}}{a^6}}}\label{t_equation}
\ee
and
\be 
\phi=\phi_0\pm
\int \frac{\sqrt{6}}{a}{\frac{\sqrt{\displaystyle\rho(a)-\frac{F}{a^6}}\,da}
{\sqrt{\displaystyle\Lambda+\kappa^2\rho(a)+
\frac{\tilde\kappa^4}{12}\rho(a)^2+\frac{{\varpi}}{a^6}}}},
\label{phi_a_hub2}
\ee
where $t_0$, $\phi_0$ are integration constants. We will  
choose the plus sign in Eq. (\ref{t_equation}) so that we have expansion as
time grows, and will also make the same sign choice in 
Eq. (\ref{phi_a_hub2}) for reasons that will become
clear in the lines below.

Our main objective at this stage is
 to find expressions for $F(a)$ so that they are suitable for depicting
a given kind of inflationary behavior, and in particular
that associated with exponential potentials. 
We assume then that the expansion is driven by 
a scalar field striving to make its potential energy as small as possible. 
If the potentials have positive slopes, 
the only possibility is that the scalar field grows with time, 
and in particular with the scale factor. Therefore, we conclude
that given our purposes we have to   
choose the plus sign in Eq. (\ref{phi_a_hub2}). 

One simple way of satisfying the requirements we just made on the potential 
is to set
\be
F(a)=f a^{n}
\ee
with $n<6$ and $f>0$. This gives
\be
V(a)=f a^{n-6}\label{pot}.
\ee

Note that, as pointed out in Ref. \cite{ShiMaeSas00}, the consequence of
the usual assumption $\tilde \Lambda<0 $ allows for an arbitrary  value of
$\Lambda$, and in particular it can be chosen so that it vanishes or takes
some other wanted value . Let us set $\Lambda=0$ for the time being, 
together with $c=0$. We then get 
\be
\frac{d\phi}{da}=\pm\frac{\sqrt{6-n}}{\sqrt{\displaystyle
a^2\kappa^2+\frac{n{\varpi}}{6fa^{n-2}}+
\frac{f\tilde\kappa^4}{2n a^{4-n}}}}\label{phi_a_hub3}.
\ee

The relation $V=V_o e ^{-k\phi}$ will only hold if 
$\phi\propto \log a$, so the question arises of whether that condition
is compatible with Eq. (\ref{phi_a_hub2}). 
In order to find the answer we power expand $d\phi(a)/da$. 
around $a=\infty$  and keep only the first term in the series. 
It can be
noticed immediately that the required condition $\phi\propto \log a$ will
only be satisfied asymptotically for large $a$ if
 $0<n<6$.  Therefore, the study will be restricted to 
just those 
cases. This would then give
\be
V\sim f e^{-\sqrt{6-n}\,\kappa\phi}\label{pot_ground},
\ee
with 
\be
\phi\sim\frac{\sqrt{6-n}}{\kappa}\log a\label{phi_ground}.
\ee

There are three exactly integrable cases, namely $n=2,3,4$,
and the corresponding potentials turn out to be functions of 
$e^{-\kappa\sqrt{6-n}\phi}$. For other values of $n$  one can only
give approximated expressions for late time limits. The
leading terms in the solution $\phi(a)$ will be 
the ${\varpi}$-dependent or the 
$\tilde\kappa$-dependent ones
depending on whether $0<n<3$ or $3<n<6$.

With the aim of giving the expressions
in a form as simple
as possible we define the coefficients
\be
s_n=\frac{n\,\varpi }{24f{{{\kappa }}}^2}
\ee
and 
\be
b_n=\frac{f{{{\tilde\kappa }}}^4}{8n{{{\kappa }}}^2}=\frac{3f}{4n\lambda },
\ee
where the letters $s$ and $b$ indicate that they are related to the shear
and the brane tension respectively. We outline the expressions for the scalar
field and potential in Table~\ref{table}.

\begin{widetext}
\begin{center}
\begin{table}[h!]
	\begin{center}
		\begin{tabular}{c|ccc}

		&$ \large\phi(a)$ & $ \large V(\phi)$\\
			\hline
			\\
			$n=2$&$\displaystyle\frac{1}{\kappa}{\log\frac{{a^2 + 2s_2 + 
       {\sqrt{a^2\left(a^2+4s_2\right)+ 4b_2 }}}}{2}}$ &$\displaystyle\frac{e^{2\kappa \phi }\,f}
  {{\left( {\left( e^{\kappa \phi } - s_2\right) }^
        2 -b_2\right) }^2}$\\
			\\
		$n=	3$&$\displaystyle\frac{2}{\sqrt{3}\,\kappa}{\log \frac{{a^{{3}/{2}} + {\sqrt{a^3 + 4(b_3 + 
            s_3)}}}}{2}}$&$\displaystyle\frac{e^{{\sqrt{3}}\,\kappa \phi }\,f}
  {{\left(  e^{{\sqrt{3}}\,\kappa \phi }- b_3  -s_3
       \right) }^2}$\\
       \\
				$n=4$&$\displaystyle\frac{1}{ {\sqrt{2}}\,\kappa }{\log \frac{a^2 + 2{b_4} + 
       {\sqrt{a^2\left(a^2 + 4b_4 \right)+ 4s_4}}}{2}}$& $\displaystyle\frac{e^{{\sqrt{2}}\,\kappa \phi }\,f}
  { \left(e^{{\sqrt{2}}\,\kappa \phi } - b_4\right)^2- s_4 }$\\
  \\
 $0<\!n\!<3$ &$\displaystyle{\frac{2\,{\sqrt{6 - n}}}{n\kappa }\log \frac{a^{{n}/{2}} + {\sqrt{a^n + 4\,s_n}}}
      {2}}$& $\displaystyle \frac{f\,e^{{\sqrt{6 - n }\, \kappa \phi }}}{{\left( \displaystyle e^{{n\kappa\,\phi }/{{\sqrt{6 - n}}}} - {s_n} \right) }^
   {\scriptstyle2\left( 6 - n \right)/n}}$\\
   \\
$3<\!n\!<6$ &$\displaystyle   {\frac{2}{{\sqrt{6 - n}}\,\kappa}\,\log \frac{{a^{3 - {n}/{2}}}+ {\sqrt{ {a^{6 - n}}+4b_n}}}{2}}$ & $
\displaystyle\frac{e^{{\sqrt{6 - n}}\,\kappa \phi }\,f}
  {{\left( e^{{\sqrt{6 - n}}\,\kappa \phi } - b_n \right) }^2}$
		\end{tabular}
	\end{center}
	\caption{Scalar fields and potentials for the integrable cases. The expressions  in the last two rows are only valid in the large $a$ approximation.}\label{table}
\end{table}
\end{center}
\end{widetext}

We have chosen the integration constant $\phi_0$ in such a way that  in the large $a$ and large $\phi$ limits we consistently recover the same expressions as in Eqs. (\ref{pot_ground}) and  (\ref{phi_ground}). Solutions similar to these
were obtained by Barrow and Saich \cite{BarSai93} in a study of Friedmann universes containing 
a perfect fluid and a massive scalar field under the requirement that 
the kinetic and potential energies of the scalar field be proportional.

Before we go further, a remark regarding
the choice of the value of $\Lambda$ is in order. We have found three exact 
solutions for the $\Lambda=0$ case (see Table 1), but we can, by the same 
token, find another three for any  non null value of $\Lambda$. 
This requires a little generalization that lies in including in the 
potential a term
acting like a cosmological constant. To this end we set
$
F(a)=f a^{n}+ h a^6$
which will obviously yield $
V(a)=f a^{n-6}+h.$ Now, for the choice  
\be
h=-2\frac{3{{{\kappa }}}^2 \pm{\sqrt{9{{{\kappa }}}^4 - 
3\Lambda{{{\tilde\kappa }}}^4}}}
  {{{{\tilde\kappa }}}^4}  
\ee
one gets 
\be
\frac{d\phi}{da}=\pm\frac{\sqrt{6-n}}
{\sqrt{\displaystyle
a^2\left(\kappa^2+{h\tilde\kappa^4}\right)+\frac{n{\varpi}}{6fa^{n-2}}+
\frac{f\tilde\kappa^4}{2n a^{4-n}}}}\label{phi/a_bis}.
\ee
This can be obtained from Eq. (\ref{phi_a_hub2}) upon the replacement 
$\kappa^2\rightarrow\kappa^2+h\tilde\kappa^4$ and the same applies for
the expression giving $dt/da$. Therefore, the results for the previous set of 
solutions can be easily extended to any $\Lambda\ne 0$ value.


\section{Kinematics and inflationary behavior}

The subject in hands now is the kinematics of our cosmological models.
Let us first see if the energetic features of our cosmologies are suitable 
for inflation. A necessary condition for the occurrence
of accelerated expansion is that the potential  part of the 
energy density dominates, and for this to be possible the potential will
have to be flat enough. Let us denote
\be
\rho_{kin}=\frac{1}{2}\dot \phi^2
\ee
and
\be\rho_{pot}=V,
\ee so that $\rho=\rho_{kin}+\rho_{pot}$. 
Specifically, we have
\be
\rho_{kin}=\frac{c}{a^6}+\frac{6-n}{n}\,\rho_{pot}.
\ee
This means that the kinetic and potential energy decrease at the same
pace at late times (large $a$) if $c\ne0$, and at any time if $c=0$.
Moreover, if the latter holds we will have a barotropic equation of state $p=(1-n/3)\rho$. 
This means we have found the exact potentials for radiation ($n=2$ , $p=\rho/3$), dust  ($n=3$, $p=0$), 
and cosmic strings ($n=4$, $p=-\rho/3$) models. Note that, in the $c\ne0$ case the energy density could be dominated by
its kinetic part at early times.

It has to be pointed out that there is one more integrable case of Eq. (\ref{phi_a_hub2}), the stiff fluid model ($f=0$), which was  found in Ref. \cite{CheHarMak01}.

Typically, the inflationary behavior depends on  the slope of the potential, 
but comparisons between potentials with different slopes must be done for 
identical values of the scalar field. In our models   we have
${\partial^2 V}/{(\partial {\varpi}\partial\phi)}\ge 0$
and 
${\partial ^2V}/{(\partial \tilde \kappa\partial\phi)}\ge 0$.
This can be checked using $\partial V/\partial\phi=(dV/da)(da/d\phi)$ and
Eqs. (\ref{pot}) and (\ref{phi_a_hub2}). This would seem to 
indicate that the  anisotropy and the
quadratic corrections would make inflation less likely. 

Nevertheless, the definitive answer is provided by the deceleration
factor $q$, which is defined through \be
q=-\frac{\ddot a a}{\dot a^2}\label{q},
\ee and can be evaluated  on calculating $t=t(a)$ from Eq. (\ref{t_equation}). In Table~2 we summarize
the behavior of $q$ at late and early times.

\begin{table}[h!]
	\begin{center}
		\begin{tabular}{r|c|c|c}
	\multicolumn{2}{c|}{} & $t-t_0$ & $q$  \\
		\cline{1-4}
		&&&\\
&$0<n<3$ & $\displaystyle\frac{n a^{6-n}}{f(6-n)\tilde \kappa^2}$&$5-n$\\
&&&\\
\cline{2-4}
&&&\\
$a\rightarrow 0\quad$&$n=3$& $\displaystyle\frac{a^3}{{\sqrt{3\varpi  + {f}^2\tilde\kappa ^4}}}$&$2$\\
&&&\\
\cline{2-4}
&&&\\
&$\quad3< n< 6\quad$ &$\displaystyle\frac{a^3}{{\sqrt{3\varpi}}}$ &$2$\\
&&&\\
\hline
& &  & \\
$a\rightarrow \infty$ &$0<n<6$& $\displaystyle\frac{{\sqrt{2 n}}\,a^{3 - {n}/{2}}}
    {\left( 6 - n \right) \,{\sqrt{f}}\,{{\kappa }}}  $& $\displaystyle \,2-\frac{n}{2}$  \\
& &  & \\
	\cline{2-4}
	\hline
		\end{tabular}
	\end{center}
	\caption{Asymptotic expressions for $t$ as a function of $a$, together with the corresponding deceleration factors.}\label{table}
\end{table}

For large enough $a$ we get $q=(4-n)/2$,
so inflation will only occur at late times if and only if $n>4$. This 
result is completely equivalent to that obtained in  \cite{AguFeiIba93}
in the framework of general relativity.
There is a persistent non inflationary behavior in the asymptotic regime
of the models if the potential is too steep. For $a\rightarrow 0$, however, 
inflation will by no means occur.

Let us push a bit further the fact that the sign changes in $q$ mark the 
transition between an inflationary
and a non inflationary epoch. As follows from Eq.
(\ref{q}) accelerated expansion  will occur if $\ddot a>0$, which in 
our models holds if
\be
3\,\left( n -4\right)n{a}^{6 + n}f
{{{\kappa }}}^2 + 
  3\,\left(n -5  \right){a}^{2n}{f}^2{\tilde\kappa }^4-2n^2{\varpi} {a}^6>0
\label{q_con}.\ee  
We see that $n>4$ is a necessary, but not sufficient, condition for
accelerated expansion, and
also a sufficient condition for $a\rightarrow \infty$, as proved above. Nevertheless,
Eq. (\ref{q_con}) has in store some interesting information that
 will highlight how peculiar inflation is  in brane cosmology.

Let us make $a_{\mathrm{ inf}}$ stand for the value of the expansion rate
above which the model expands.  We will now study, by numerical techniques, how
$a_{\mathrm{ inf}}$  depends on the parameters of the model. First, we will focus on the
dependence on the equation of state. After that, we will turn to consider
the effect of the magnitude of the anisotropy and the brane tension.

Numerical calculations indicate that, in the relativistic limit, as is well known, the larger $n$, the  smaller $a_{\mathrm{ inf}}$  (i.e. the earlier inflation begins), whatever the value of $\varpi$. Inspection suggests, however, that this monotonic behaviour may not hold for the brane case, because the
term in Eq. (\ref{q_con}) associated with the extra dimensional corrections
changes sign at $n=5$. Moreover, the values of $a_{\mathrm{ inf}}$  in the vicinity of $n=5$
must be very much alike the values corresponding to the relativistic limit (because
the extra dimensional corrections will be negligible). Numeric evaluations show that
for $n$ close to $4$ the behaviour of $a_{\mathrm{ inf}}$  is almost identical in the brane
and the relativistic case. However, as $n=5$ is approached, the monotonic decrease
ceases and $a_{\mathrm{ inf}}$  begins to grow with $n$, till it reaches a maximum value
before $n=5$. Note that the value of the maximum of $a_{\mathrm{ inf}}$  does not depend on
$\varpi$. From that point on, there is again monotonic decrease, but is more pronounced that in the relativistic case. Summarizing, the conditions for inflation are rather different in the brane and relativistic setups. Nevertheless, several restrictions on the parameters of the models have to be considered. On the one hand the 
potential must
remain much smaller than the fourth power of the four-dimensional 
Planck's mass $M_P$
so that classical physics is valid. On the other hand, the five-dimensional 
Planck's mass $\tilde M_P$ 
is typically much smaller than its four-dimensional counterpart. 
In Fig. 1
we have  represented $\log a_{\tiny\hbox{inf}}$ as a function of $n$, 
for several values of ${\varpi}$, within the ranges $4<n<5$ and $1/2\le\varpi\le 8$
respectively, 
choosing $\tilde M_P$ considerably smaller than $M_P$.

\begin{figure*}
\label{figure:t(a)curves}\centering
\includegraphics[width=.49\textwidth]{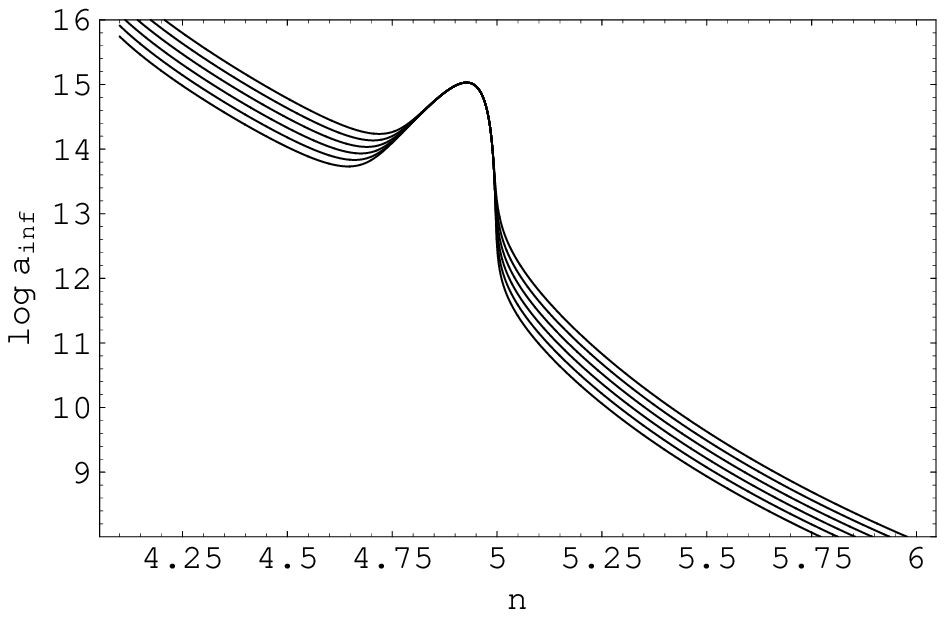}\includegraphics[width=.49
\textwidth]{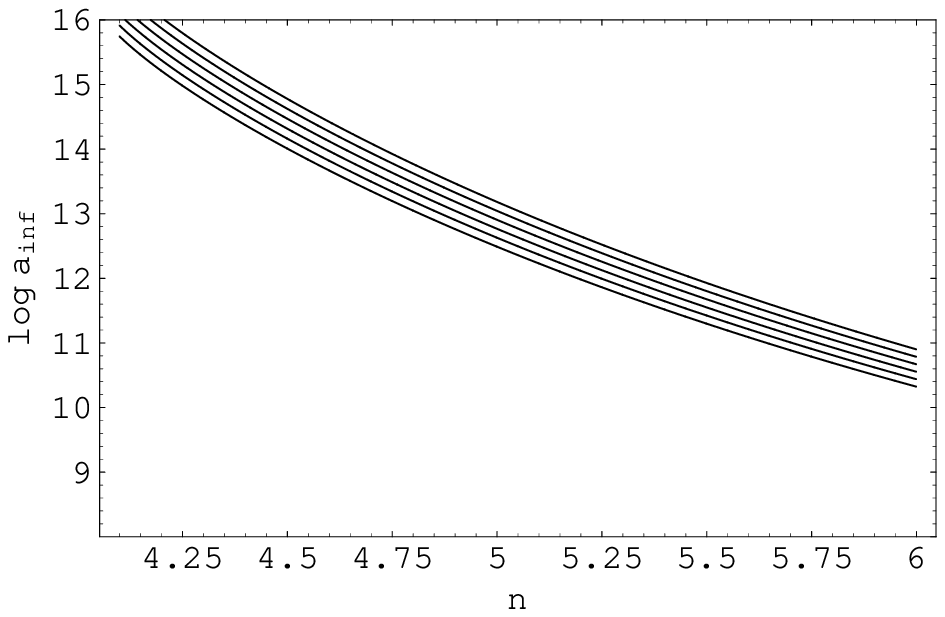}
\caption{Plot of $\,\log a_{\hbox{\tiny inf}}$ as a function of $n$ for ${\varpi}=0.25,0.5,1,
2,4,8$, $a(0)$=1, $f=10 ^{-30}$,  
$\kappa=8\pi$, and $\tilde\kappa=8\pi\times 10^9$. 
for the figure on the left
and $\tilde\kappa=0$ for the  figure on the right.
Here the  Planck mass on the brane
$M_P$ has been set to one.
Lower curves in each plot correspond to lower values of 
${\varpi}$.}
\end{figure*}

We have seen, also by numeric means, that $a_{\mathrm{ inf}}$  increases with
the shear and decreases with the brane tension\footnote{The shear increases with $\varpi$, but  the brane tension decreases as $\tilde
\kappa$ increases.}.  However, those very changes  in the parameters
which act to delay inflation,   make  the Hubble factor grow in turn \cite{Maa00}, and inflation can be sustained by potentials that would  not be able to do so in a relativistic setup \cite{CopLidLid01}. Thus, both the anisotropies
and the extra dimensional corrections seem to act locally 
in a way conducive to inflation. In order to determine whether this is
also the case in a global way, it would be necessary to evaluate the
number of e-foldings between the beginning and the end of inflation. So, 
we would have to find physical arguments allowing to know when inflation 
ends, and then calculate the amount of inflation achieved while accelerated
expansion persists.

Unlike what happens with  the time at which inflation ends, we can 
determine for our models the time at which inflation begins.
 We have seen that  the larger
${\varpi}$ or $\tilde\kappa$, the larger $a_{\mathrm{ inf}}$ . But because of the
way the Hubble factor behaves, the value of $\Delta t$ associated
with a given value of $a$ grows too. Due to the lack of exact expressions
for the $t(a)$ curves for the inflationary cases ($n>4$) one can only 
find the corresponding value of time
by using approximate solutions or  numerical
integration; we will choose the second method. In Fig. 2 
we have  plotted the $t(a)$ 
curves for of $n=4.5$ and for the same values of $\tilde M_P$ and ${\varpi}$ as in Fig. 1;
we have also indicated in the picture the value of $a_{\mathrm{ inf}}$ as a function
of time.

\begin{figure}[h!]
\label{figure:t(a)curves}
\begin{center}
\includegraphics[width=8cm,height=5.3cm]{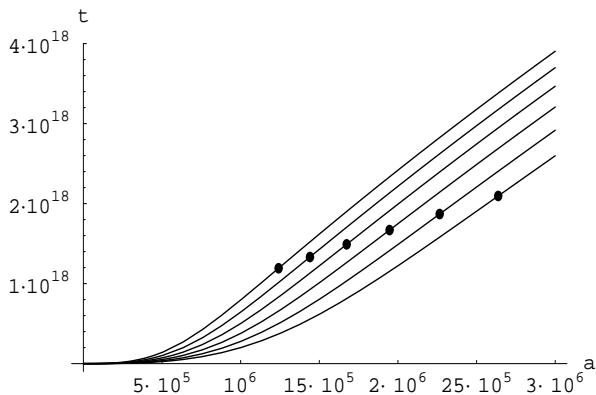}
\end{center}
\caption{Plot of the $t(a)$ curves for ${\varpi}=0.25,0.5,1,
2,
4,8$, $a(0)$=1, $f=10 ^{-30}$, $n=4.5$, $f=10^{-30}$,
$\kappa=8\pi$ and $\tilde\kappa=8\pi\times 10^9$. Here the  Planck mass on the brane
$M_P$ has been set to one.
Lower curves correspond to higher values of 
${\varpi}$. Each dot marks the beginning of inflation}
\end{figure}

For the sake of completeness and comparison,  let us investigate whether our models isotropize or not.
We have seen above that the influence of the anisotropies in the inflationary
behavior becomes less and less important as time goes by.
The anisotropic shear parameter, which is defined as 
\be
\Omega_{\mathrm {shear}}=	\frac{\sigma^2}{6H^2}=\frac{\sum_{i=1}^3(H_i-H)^2}{6H^2},
\ee 
reflects that too. Using Eqs. (\ref{P})-(\ref{varpi}), and after a little algebra, 
we get 
\be
\Omega_{\mathrm {shear}}=\frac{{\varpi} P^2}{3 H^2},
\ee 
and according to Eq. (\ref{phi_a_hub2}) one finally has
\be
\Omega_{\mathrm {shear}}=\frac{{\varpi} a ^6 n^2}
{6\left(2a ^{6+n}n f\kappa^2+3a^{2n}f^2\tilde\kappa^4
+\displaystyle{a^6n^2{\varpi}}\right)}.
\ee
In the inflationary cases, since  $n>4$, we see that $\Omega_{shear}$ decreases  monotonically with time, tending to a null value.
Moreover, the larger $n$ and $\tilde \kappa$ the more
rapidly it decreases. Thus, although we have genuinely anisotropic models
at early times, anisotropy gets completely dissipated in the course of the evolution. So, our 
models do isotropize at late times, and both faster inflation and extra 
dimensional effects make the isotropization process more efficient. 

\section{Generalization to multi-field models}
Let us now take advantage of the results in the previous sections to discuss how 
the situation would change if we had multiple fields.  Assisted inflation \cite{LidMazSch98}, in its simplest 
form, is realized with configurations of identical multiple fields with no interaction among themselves, but which self-interact through the same potential (see also \cite{ColHoo00, HooFil00}). Aspects of this proposal in connection with the brane scenario have been discussed in \cite{MenLid00,Maj01,Pia02}.

 If we now make $\phi$ denote
each scalar field and $V(\phi)$ each individual potential, we can obtain the multifield
version of our models just by making the replacement
\begin{eqnarray}
\kappa\rightarrow \sqrt{N}\kappa,\\
\tilde \kappa\rightarrow \sqrt{N} \tilde\kappa,
\end{eqnarray}
where $N$ denotes the number of fields. It can be immediately seen than the larger $N$,
the larger the Hubble factor, so that the assisted inflation proposal is valid in the brane scenario
too.

The asymptotic individual potentials will become
\begin{equation}
V\sim e^{-\kappa\sqrt{N(6-n)}\phi}.
\end{equation}
Thus, an increase in the number of fields makes the potentials turn flatter,  which means that late time inflation becomes likelier.

The  condition for having $q<0$ reads now 
\be
3\left( n -4\right)n\,{a}^{6 + n}f
\,{{{\kappa }}}^2 N + 
  3\left(n -5  \right){a}^{2n}{f}^2{\tilde\kappa }^4 N^2-2n^2{\varpi} {a}^6>0
\label{q_con2},\ee  
so, the larger $N$, the less anisotropy influences the occurrence of inflation. 
This is also reflected by the fact that $\Omega_{\mathrm{shear}}$ decreases with
growing $N$.
 Nevertheless $n>4$ is still necessary for inflation, no matter how large $N$ is. 

The behaviour of  $a_{\mathrm{inf}}$ as a function of
$n$ will change when $N$ grows. We have seen in the previous section that, unlike
in the relativistic case, $a_{\mathrm{inf}}$ is not a monotonically decreasing function of $n$, and that
it has a maximum near $n=5$. It is not difficult to see that a larger $N$ 
increases the  value of $a_{\mathrm{inf}}$ at the maximum and 
makes wider the range of $n$ in which   $a_{\mathrm{inf}}$ grows with growing $n$.
Summarizing, the addition of fields accentuates the differences between relativistic
and brane cosmologies.

\section{Conclusions and future prospects}
We have studied Bianchi I type brane cosmologies with 
barotropic perfect fluids, and we have shown that they can be interpreted 
in the language of scalar fields as exponential potential models. From the expressions
here obtained, it follows that the models behave, at late times, as if the potential had only one exponential term. Consistently, the differences between the brane and relativistic models get  washed off as the models evolve in time. The effects of the anisotropy disappear in a similar fashion as well.

Our main focus point has been the average evolution of the models, 
in comparison to their relativistic counterparts. In particular, we have studied
the features of the inflationary behaviour that can be traced back 
to extra dimensional effects but not to the bulk.

It is a well-known fact that, in the  relativistic context,  
the condition $3p <-\rho$ (which corresponds to $n>4$) is necessary and sufficient for 
inflation to occur. Expressed in terms
of the potential, it translates into the requirement that the slope of the asymptotically dominating potential be smaller than ${\sqrt 2}\kappa$. In brane cosmology, that condition remains necessary for inflation, although we show that it is not sufficient. 

Using numeric means we have found some other new interesting results
in connection with the kinematics of the models. In the
relativistic case, the  value of the average expansion factor at 
the beginning of inflation decreases  monotonically as the ratio $p /\rho$ becomes more negative. In the brane models,
the behaviour is basically the same, except for the vicinity of  
$3p <-2\rho$, where the monotonic decrease turns into monotonic increase, until 
a local maximum in the function is reached. Moreover, the value of the expansion
rate at that extreme point does not depend on the amount of anisotropy.

Taking into account the tight restrictions to the parameters of the models set by physical arguments, our  calculations show that the effect of the anisotropy in the evolution will only be noticeable if inflation begins very late, 
which is not a desirable situation. 

We have discussed a possible realization of
the assisted inflation proposal, which is underepresented
in the literature on brane cosmology.  In particular, we have considered
generalizations to multi-field configurations of non-interacting
scalar fields. The conclusion reached is that 
the aforementioned requirement on  the slope of 
the potential still applies. On the other hand,  an increase in the number of fields $N$
will make the Hubble factor grow, so that   inflation will be likelier
This means that the brane version of the 
assisted inflation proposal is also feasible. Moreover, as $N$ grows
the models become more isotropic. 

A possible extension to this work would be exploring the possibility of generalizing these solutions to include bulk effects. It would also be interesting to carry out similar studies for anisotropic spacetimes with non-trivial space-curvature. Bianchi V type models would be the obvious candidate for a first attempt because they are the simplest 
generalization of the open FRW models.

\section*{Acknowledgements}
We are grateful to L.P. Chimento, S. Jhingan and M.A. V\'azquez-Mozo for discussions
and suggestions. 
This work was supported by the Spanish Ministry of Science and Technology
jointly with FEDER funds through research grant  BFM2001-0988,
and the University of the Basque Country through research grant 
UPV00172.310-14456/2002. Ruth Lazkoz's
work is also supported by the Basque Government through fellowship BFI01.412.

\end{document}